\documentclass[12pt]{article}
\title{Comment on ``Stress induction in the bacteria {\it Shewanella
oneidensis\/} and {\it Deinococcus radiodurans\/} in response to
below-background ionizing radiation'', Castillo, {\it et al.\/} Int. J. Rad.
Biol., 2015; Early Online DOI: 10.3109/09553002.2015.1062571}
\author{J. I. Katz \\
Department of Physics\\
McDonnell Center for the Space Sciences \\
Washington University, St. Louis, Mo. 63130 USA \\
{\tt katz@wuphys.wustl.edu}}
\begin{document}
\maketitle
Castillo, {\it et al.\/} \cite{C15} report that reducing levels of ionizing
radiation below background inhibits growth in these two species of bacteria.
This result has the remarkable implication that background levels of
radiation, at which there is a very small probability of even one ionizing
event in a bacterium in its replication time, are sufficient to facilitate
its growth.

In this study natural backgrounds were simulated by a gamma-ray irradiation
of approximately 100 nGy h$^{-1}$.  For bacteria with a generation time of 3
hours the energy deposition was about $3 \times 10^{-7}$ J kg$^{-1}$ per
generation.  The mean energy deposition per ionization is about 30 eV ($5
\times 10^{-18}$ J), so that simulated background corresponds to about $6
\times 10^{10}$ ionizations per kg per generation.

The mass of a {\it S.~oneidensis\/} bacterium is about $6 \times 10^{-16}$
kg, while that of a {\it D.~Radiodurans\/} bacterium is about $8 \times
10^{-15}$ kg.  Hence the number of ionizations per {\it S.~oneidensis\/} per
generation is about $4 \times 10^{-5}$ while that per {\it D.~radiodurans\/}
per generation is about $5 \times 10^{-4}$.  Only tiny fractions of the
bacteria experience even a single ionization during their lifetimes.  The
actual fractions are even smaller than these numbers indicate because
ionizations are clustered along the recoil paths of electrons that Compton
scatter the irradiating gamma rays.  Hormesis cannot be acting through a
response to ionization during the lifetime of a bacterium.

Perhaps a single bacterium lucky enough to have an ionization event produces
a super-vigorous clone that dominates its colony.  This would imply an
incubation time for the clone to dominate its colony and increase the
measured replication rate; no such incubation time is evident in the data.

If the reported result is valid, it must be produced by exposure of the
growth medium to background radiation long (thousands of hours) before
it is incorporated into the bacteria.  It would be necessary that the
hormetic effect be effective in a large fraction of the bacteria growing in
that medium; most bacterial cells must contain at least one molecule
produced by ionization.  Some free radical or other species produced by
ionizing radiation must have long enough lifetime to accumulate to
concentrations of at least one per bacterium.

This hypothesis could be tested by lengthy (thousands of hours) exposure of
the growth medium to background radiation prior to incubating bacteria in a
low radiation environment, and {\it vice versa\/}; it predicts that long
prior exposure would be hormetic, but not exposure during the brief period of
bacterial growth.  Experiments of this type could be conducted at higher
radiation intensities (and with shorter durations) to determine if hormesis
is prodeuced by irradiation during bacterial growth or if radiation
pre-conditions the growth medium.

Mammalian cells are much bigger than bacteria.  Further, a long-lived
multicellular organism might be affected by events in a small fraction of
its cells.  These arguments are therefore not applicable to the possibility
of hormesis in mammals, either at background radiation levels or at higher
levels.

\end{document}